\documentstyle[aps]{revtex}
\begin{document}
\title{Antibranes and crossing symmetry}
\author{Vipul Periwal}
\address{Department of Physics,
Princeton University,
Princeton, New Jersey 08544}

\def\dd{\hbox{d}}
\def\tr{\hbox{tr}}\def\Tr{\hbox{Tr}}
\maketitle
\begin{abstract}  Crossing symmetry appears in
  Dbrane--anti-Dbrane dynamics in the form of an analytic 
continuation from U$(N)$ for $N$ brane amplitudes to U$(N-p,p)$ for
the interactions of $N-p$ branes with $p$ anti-branes.  I consider the 
consequences for supersymmetry and  brane--anti-brane forces.
\end{abstract} 

\def\al{\alpha}
\def\be{\beta}
\def\la{\lambda}
\def\eps{\epsilon}
\def\sig{\sigma}
\def\la{\lambda}
\def\ga{\gamma}
\def\half{\hbox{$1\over 2$}}
\def\quart{\hbox{$1\over 4$}}
\def\ee{\hbox{e}}
\def\part{\partial}
In Dbrane interactions, the 
BPS property allows one to study the interactions of $N$ brane 
configurations with some confidence, but one has no reliable manner 
in which to compute the interactions of Dbranes and anti-Dbranes. This 
entails an inability to answer  questions such as unitarity 
in black hole physics.    This 
has been evident since the pioneering work of M. Green\cite{green}, which first
exhibited the supersymmetric cancellation of forces between Dbranes, 
as well as the pathologies in a string description of 
Dbrane--anti-Dbrane interactions.  
The supersymmetric cancellation was interpreted by 
Polchinski\cite{pol} as 
showing that Dbranes are  BPS states that carry Ramond charge, 
a finding which provides strong evidence 
for the existence of dualities between different string 
theories\cite{polwit}.

An interpretation of the  
pathologies in the string description of Dbrane--anti-Dbrane interactions 
was given by Banks and Susskind\cite{bs}.   While the 
cancellation of forces found for the Dbrane--Dbrane interaction is 
evidence in perturbation theory of the  coupling--constant 
independent BPS 
property, having found a non-zero force between a Dbrane and an anti--Dbrane
at closed--string tree--level, one can say very little   about the true
physics of the system since there is no reason to think that higher 
orders in  string perturbation theory will not change the physics
drastically.  An example of such a phenomenon is Strominger's 
interpretation of singularities in moduli spaces of string 
vacua\cite{andy}.  One should, therefore, attempt to 
apply general principles to resolve 
the question of Dbrane--anti-Dbrane interactions.

Clearly, particle--anti-particle 
processes are different from particle--particle processes, but   
in field theory, locality, analyticity, and CPT invariance allow the 
derivation of remarkable  properties such as crossing 
symmetry, relating particle amplitudes at physical momenta 
to anti-particle amplitudes at unphysical 
momenta, and vice versa.  We have no precise idea of the 
meaning of 
locality in Dbrane physics, but we do have a clear idea of what 
co\"ordinates are---they are the dynamical variables in a matrix 
theory that describes Dbrane interactions\cite{ed}.   A natural 
question, then, is: 
What natural holomorphy properties can be associated with these
dynamical variables, and what physical statements can one deduce
from appropriate analytic continuations?  The only guiding principles
we allow are Lorentz invariance and supersymmetry.  

Following Witten\cite{ed},
we will   consider supersymmetric Yang-Mills theory for a compact 
simple Lie group in 
$9+1$ dimensions (10dsym), reduced to a point.  The first
simple observation we make is that if one {\it holomorphically} continues
the gauge potential, $A,$ and the Majorana-Weyl fermion, $\la,$ the
`action' (now no longer real) is still invariant under the supersymmetry
transformations.  In other words, if one complexifies the gauge group,
supersymmetry still holds.  In particular,  there is a supersymmetric
gauge theory for any real form of the complexification of the compact
simple Lie algebra that we started with.

Now, consider $su(N).$ The complexification is $sl(N,{\bf C}).$ 
Possible real forms are $su(N-p,p)$ for any $p\ge 0, p\le N.$  (There 
are other real forms, such as $sl(N,{\bf R}),$ but I have not found 
a physical interpretation for these as yet.) $su(N-p,p)$ can be defined as
the Lie algebra of $N\times N$ matrices that satisfy
\begin{eqnarray}
X^{\dagger}I_{N-p,p}=I_{N-p,p}X ,
\label{define}
\end{eqnarray}
where $I_{N-p,p}$ is a diagonal matrix with $(N-p)$ 1s and $p (-1)$s on
the diagonal.
There is an obvious natural embedding of $su(N-p)$ and $su(p)$ into 
$su(N-p,p).$ This implies, in particular, that  if we interpret the
$X$ matrices as describing the interactions of $N-p$ Dbranes with 
$p$ anti-Dbranes, then the off-diagonal elements governing the
interactions of Dbranes with Dbranes, and
the interactions of anti-Dbranes with anti-Dbranes, are related by 
hermiticity, just as they 
are in the original $su(N)$ theory.  The only interactions that change
are the blocks governing the
interactions of Dbranes with anti-Dbranes, which are now related 
by $X_{ij}= - \bar X_{ji}, i\le N, j>N.$  This is 
what one might expect  for Dbrane--anti-Dbrane 
interactions.  
The intuition behind this is that closed string exchange is being modelled in
the off-diagonal elements as a product of two open string exchanges,
as is standard in representing closed string amplitudes\cite{rob}.  If
closed string exchange between like particles is modelled by 
$X_{ij}\bar X_{ij}\equiv X_{ij}X_{ji}, i,j>N \hbox{or}\ i,j \le N,$ then
closed string exchange between particles with opposite charges might  be
obtained from $X_{ij}X_{ji} = - X_{ij}\bar X_{ij}, i\le N, j>N,$ which
is enforced by eq.~\ref{define}.

I will now  show that this is exactly what happens in
the interaction of an Dinstanton and an anti-Dinstanton, following 
recent work of Green and Gutperle\cite{gg}.  In \cite{gg}, an
explicit computation of the $u(2)$ matrix model integral is performed,
precisely for the purpose of analyzing Dinstanton interactions.  The 
10dsym theory is reduced to a point, following \cite{ed}, leaving a 
model of interacting matrices  with Lorentz invariance, supersymmetry 
and a `gauge' invariance.  I will be brief because the discussion in 
\cite{gg} is  explicit, and there is nothing new involved in the
computation for $su(1,1).$  My calculation differs slightly from 
\cite{gg} in that I will not continue to Euclidean space. The 
Dbrane--anti-Dbrane integral does not admit a trivial Euclidean 
continuation, since the system is `unstable'.  

The dimensionally reduced action is 
($a,b = 0,\ldots,9$, $\eta=\hbox{diag}(+,-\dots,-),$
$\la=\la^{c}=C\bar\la^{T}=-\Gamma_{11}\la$, 
$A_{a}$ Hermitian and traceless, $F_{ab}\equiv [A_{a},A_{b}]$)
\begin{eqnarray}
S\equiv \tr\left( \hbox{$1\over 4$} F^{ab}F_{ab} - \hbox{$1\over 2$} \bar\la
\Gamma^{a}[A_{a},\la]\right).\label{pt-act}
\end{eqnarray}
The supersymmetry transformations are
\begin{eqnarray*}
\delta_{\eps} A_{a} = i\bar\epsilon\Gamma_{a} \la,\qquad \delta\la = 
\hbox{$i\over2$}\Gamma^{a}
\Gamma^{b}F_{ab}\eps, \qquad \Gamma_{11}\epsilon=-\epsilon,
\end{eqnarray*}
and gauge transformations are 
$\delta_{\al}\phi = -i[\al,\phi]$ {for}  $ \phi 
=A,\la.$ The commutator of two supersymmetry transformations is
$[\delta_{1},\delta_{2}] = \delta_{\al},$ with $\al\equiv 
-2i\bar\eps_{1}\Gamma^{a}\eps_{2}A_{a}.$  This implies the following
identification: $P_{a} \equiv i{\partial/{\partial x^{a}}}
\leftrightarrow -A_{a}.$ 

I now relax the hermiticity requirement on $A,\la,$  and consider
$A,\la$ to take values in either $su(2)$ or $su(1,1).$  In either 
case, the bosonic part of the 
action can be explicitly written as a Lorentz-invariant 
function of 3 vectors, corresponding to the three generators of either 
algebra.  I suppose $A_{3}$ is time-like, so by Lorentz invariance we
may take $A_{3}^{0}=T, A_{3}^{i}=0, i=1,\cdots,9.$   
We want to 
evaluate the function $V$ defined by 
\begin{eqnarray}
\int \prod_{i=1}^{3}\dd^{10}A_{i}\dd\la_{i} 
\ee^{iS/g^{2}} \equiv \int \dd^{10}A_{3}\dd^{10}\la_{3}
\ee^{iV(T)}.
\label{defin-V}
\end{eqnarray}
Note that I will ignore the fermionic  modes $\la_{3}$---for a 
careful discussion, see \cite{gg}.  The fermionic integrals are the 
same for either algebra, owing to the even number of fermions.
  The bosonic part of the action differs only
in a sign in the term that comes from 
$\tr\left(-\half \sum_{i=1}^{9}[A_{0},A_{i}][A_{0},A_{i}]\right) $ since
this is the only term where the anti-Hermitian generators of $su(1,1)$
appear quadratically.  Thus, in the final result for $V,$ I find (up 
to irrelevant constants with no $T$ dependence)
\begin{eqnarray}
\ee^{iV(T)}= g^{-14}\int_{0}^{\pi}\dd\theta \int_{0}^{\infty}\dd 
x\dd y (xy\sin^{2}\theta)^{3}\ee^{\pm i(x+y)}\ee^{-i 
g^{2}xy\sin^{2}\theta/T^{4} },
\label{int-V}
\end{eqnarray}
with the positive sign for $su(2)$ and the negative sign for $su(1,1).$
The integrals may be performed with an $i\eps$ prescription to give 
\begin{eqnarray}
\ee^{iV(T)}= g^{-14}\int_{0}^{\infty}\dd y {y^{3}\over{\big(1\mp 
g^{2}T^{-4}y\pm i\eps\big)^{7/2}}}\ee^{\pm iy}.
\end{eqnarray}
The upper sign corresponds to $su(2),$ in agreement with \cite{gg} if 
one
takes into account that Green and Gutperle worked with a Euclidean 
metric.  Thus, comparing to \cite{gg}, we see that the $su(1,1)$ case
has a space-like singularity, in agreement with earlier work of 
Green\cite{green}. 
The results for the two algebras are related in a very simple way:
\begin{eqnarray}
\ee^{iV}(su(2), g^{2}) \equiv f(g^{2}) \Rightarrow 
\ee^{ i  V}(su(1,1),  g^{2}) = \bar f(-g^{2}).
\end{eqnarray}

It is claimed in \cite{v} that 
eq.~\ref{pt-act} is a particular case of 
a $10+2$ dimensional generalization  based on the Nishino-Sezgin 
equations\cite{nish}, which are 
the fundamental equations  underlying string theory.  \cite{v} was
 an attempt to find a background independent 
Lorentz-covariant theory, motivated by \cite{town,banks}. The form of
the supersymmetry algebra, the fermion equation of motion and the 
comparison to \cite{ed} suggested that these equations were
T-dual.  Since the 
appearance of \cite{v}, Ishibashi, Kawai, Kitazawa and 
Tsuchiya\cite{bashi} have shown directly that the type IIB string is
related to eq.~\ref{pt-act}.  (Apparently these authors were  aware of
\cite{v}, but chose not to refer to this work\cite{kita}.
Compare especially their comments on the relation to 
\cite{banks}, and on T-duality.)  If the claim in \cite{v} is correct,
the instanton action\cite{ed}, eq.~\ref{pt-act} 
(or, for the purposes of F theory\cite{blen}, the fully dimensionally
reduced Nishino-Sezgin equations\cite{nish}), is the object of 
fundamental interest.  However, if we restrict  to just $su(N)$ 
matrices, we are  restricted to only describing Dinstanton processes.  

While \cite{banks} gave a light-cone parton interpretation to 
D0brane quantum mechanics\cite{ed}, motivated by \cite{dewit,town},
and therefore did not need to consider anti-D0branes, we are 
considering a Lorentz-invariant `parton' model (abusing terminology,
since the fundamental objects in \cite{v} are not `dynamical', being
generalizations of Dinstantons!). (For a clear discussion of the 
difference between these types of parton models, see Thorn\cite{thorn}.)
If the theory of matrices on a point\cite{v} (point theory) is to 
describe all physics, it must admit Dinstanton--anti-Dinstanton
interactions naturally.  I have attempted to show in
the present work that point theory does, indeed,  naturally include such
interactions.  

In a broader context,
as has been noted since the early days of general relativity and the
quantum theory\cite{wigner}, an operational definition of an event 
is fraught with difficulty.
In an intuitive sense, the D-instantons are `events', and the 
off-diagonal matrix elements are the connections between events. 
Compare this to
general relativity, where one has local Minkowski co\"ordinates about
each event, with gravity providing the matching between these local
co\"ordinate systems.    
The approximation in which one replaces $A_{0}$ by a continuum
time-translation generator is only valid when velocities are small.  
 Large velocities can only be measured by
frequent measurements, in other words, by considering events 
separated by small time-steps.  Therefore, by Lorentz invariance, 
these measurements 
must necessarily involve the complete non-commuting structure of the
time-translation operator, just as short distance physics 
does\cite{ed}.  This comment should be related  to the
work of Sathiapalan\cite{sath}, in light of the issue of massive 
modes and nonlinear effective actions such as the Dirac-Born-Infeld 
action\cite{vv}.   Finally, if the nature of time as a continuous
variable must be evaluated afresh for large velocities,  
issues such as unitarity in black hole evaporation must be 
appropriately rephrased, and may not be susceptible to simple-minded 
Hamiltonian analyses.

\acknowledgements

I am grateful to I.~Klebanov, S.~Mathur and L.~Thorlacius for
helpful conversations.
This work was supported in part by NSF grant PHY96-00258.

\end{document}